\documentclass[fleqn,12pt,twoside]{article}
\usepackage{espcrc1}
\usepackage{epsfig}

\newcommand{\AmS}{{\protect\the\textfont2
  A\kern-.1667em\lower.5ex\hbox{M}\kern-.125emS}}

\title{ Elliptic Flow Measurements with the PHENIX Detector }
\author{ Roy A. Lacey\address[USBCHEM]{Chem. Dept., SUNY Stony Brook }  for the PHENIX Collaboration\thanks{For the full PHENIX Collaboration author list and acknowledgements see the contribution by W.A. Zajc (K. Adcox {\it et al.}) in this volume.}}

\begin{document}

\maketitle

\normalsize

\begin{abstract}
             {\small Two particle azimuthal correlation functions are presented 
for charged hadrons produced in Au~+~Au collisions at RHIC ($\sqrt{s_{_{NN}}}=130$~GeV).
The measurements allow for the determination of elliptic flow without event-by-event estimation of the reaction plane. The measured correlation functions indicate elliptic flow values ($v_2$) which show significant sensitivity to both the collision centrality and the transverse momenta of emitted hadrons.}
\end{abstract}

\section{Introduction}

	One of the important goals of current Relativistic Heavy Ion reaction 
studies is the creation and study of nuclear matter at high energy 
densities\cite{phenix_1,star_flow2000,phobos1,lastcall-qm99}.
Central to these studies are open questions related to the properties
of high-energy-density nuclear matter, and whether or not conditions 
for the predicted transition to its quark gluon plasma (QGP)
phase are achieved. Such a phase of deconfined quarks and gluons has been predicted
to survive for $\approx 3-10$ fm/c in Au~+~Au collisions\cite{zhang99} 
at the Relativistic Heavy Ion Collider (RHIC), and several experimental 
probes have been proposed for its possible detection 
and study\cite{lastcall-qm99}. Elliptic flow is one such probe that has attracted considerable recent attention\cite{olli92,sor97,dan98,pin99,teaney2001,kolb2001}.

\section{Elliptic Flow}

	Over a broad range of beam energies, elliptic flow can be attributed to a delicate balance between (i) the ability of a fast pressure build-up to generate a rapid transverse
expansion of nuclear matter and (ii) the passage time
for removal of the shadowing effect on participant hadrons by the
projectile and target spectators\cite{sor97,dan98}.
If the passage time is long compared to the expansion time, spectator
nucleons serve to block the path of participant hadrons emitted toward
the reaction plane, and nuclear matter is squeezed-out perpendicular to 
this plane giving rise to negative elliptic flow.  For shorter passage 
times, the blocking of participant matter is significantly reduced and 
preferential in-plane emission or positive elliptic flow is favored because 
the geometry of the participant region exposes a~larger surface area in the 
direction of the reaction plane. Thus, elliptic flow is predicted and found 
to be negative for beam energies $< 4$~AGeV and positive for 
beam energies $ > 4$~AGeV \cite{sor97,dan98,pin99}. For RHIC energies, strong Lorentz 
contraction and very short passage times lead to significant reduction
of shadowing effects, and positive elliptic flow is expected\cite{olli92,teaney2001,kolb2001}.

\subsection{Azimuthal Correlation Functions}

	In this contribution, two-particle azimuthal correlation 
measurements are exploited to evaluate the elliptic flow of charged hadrons emitted in Au~+~Au collisions ($\sqrt{s_{_{NN}}}=130$~GeV) at RHIC. There are several important benefits which are afforded by these measurements. First, they circumvent the need for full azimuthal acceptance. Second,  they allow the determination of elliptic flow 
without event-by-event estimation of the reaction plane and the associated 
corrections for it's dispersion. Thirdly, these correlation measurements can serve 
to minimize many important systematic uncertainties (detector 
acceptance, efficiency, etc) which can influence the accuracy of elliptic flow measurements\cite{lacey93}.
	
 	The colliding Au beams ($\sqrt{s_{_{NN}}}=130$~GeV) used in these
measurements have been provided by the Relativistic Heavy Ion Collider 
at Brookhaven National Laboratory (BNL). Charged reaction products were 
detected in the east and west central arms of the PHENIX 
detector\cite{phenix_1,phenix_qm01}. Each of these arms subtends 90$^{o}$ 
in azimuth $\phi$, and $\pm 0.35$ units of pseudo-rapidity $\eta$. 
The axial magnetic field of PHENIX (0.5 Tesla) allowed for the tracking of 
particles with $p_t > 200$ Mev/c in the fiducial volume of both arms. 
The Zero Degree Calorimeters (ZDC), were used in conjunction with the 
Beam-Beam Counters (BBC), to provide a trigger for a wide range of 
centrality (cent) selections\cite{phenix_qm01}. 

         The present data analysis is based on the use of the second 
Fourier coefficient, ${\lambda}_2 = \langle \cos{(2 \Delta\phi}) \rangle $, 
to measure the anisotropy of the distribution 
in the azimuthal angle difference ($\Delta\phi = \phi_1 - \phi_2$) 
between pairs of charged hadrons\cite{lacey93};
\begin{equation}
 {dN \over d{(\Delta\phi})} \propto \left[ 1 + 2\,{\lambda}_1\cos(\Delta\phi)
+2\, {\lambda}_2\cos(2\Delta\phi)  \right].
  \label{dist}
\end{equation}
 The value of $\sqrt{{\lambda}_2}$ can be identified with the second Fourier 
coefficient $v_2$, commonly used to quantify elliptic flow with respect to 
the reaction plane\cite{Postkanzer98}.
 
 Following an approach commonly exploited in interferometry
studies, a two-particle azimuthal correlation function can be defined as follows\cite{lacey93,wang91}

\begin{equation}
 C(\Delta\phi) = {N_{cor}(\Delta\phi)\over{N_{uncor}(\Delta\phi)}}, 
  \label{ratio}
\end{equation}
where $N_{cor}(\Delta\phi)$ is the observed $\Delta\phi$ distribution
for charged particle or track pairs selected from the same event,  
and $N_{uncor}(\Delta\phi)$
is the $\Delta\phi$ distribution for particle pairs selected from
mixed events. Events were selected with a collision vertex position, 
$-20 <$~z~$< +20$~cm, along the beam axis.
Mixed events were obtained by randomly selecting each member 
of a particle pair from different events having similar centrality 
and vertex position. Event centralities were obtained by a series of 
cuts in the space of BBC versus ZDC analog response\cite{phenix_1}; they reflect 
percentile selections of the total interaction cross section 
of 7.2 barns\cite{phenix_1,phenix_qm01}.

\begin{figure}[htb]
\begin{minipage}[t]{16cm}
\epsfig{file=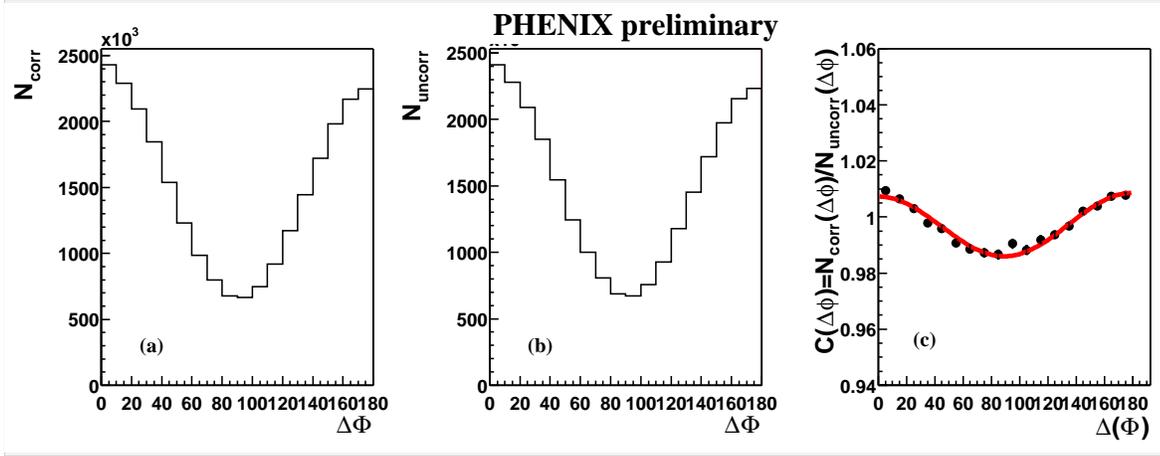,width=15.5cm}
\vskip -.75cm
\caption{\small \em   
	$\Delta\phi$ distributions for charged particle or track pairs 
selected from the same event (a) and for particle pairs selected from 
mixed events (b). The resulting correlation function $C(\Delta\phi)$, 
obtained from the ratio of these two distributions is shown in panel (c). 
}
\label{fig:cor_fun}
\end{minipage}
\end{figure}

        Figs.~\ref{fig:cor_fun}a~and~b show representative distributions 
for particle pairs obtained from the same event ($N_{cor}(\Delta\phi)$) and 
those obtained from mixed events ($N_{uncor}(\Delta\phi)$) respectively.
The correlation function $C(\Delta\phi)$, obtained from the ratio of these two 
distributions (cf Eq.~\ref{ratio}) is shown in Fig.~\ref{fig:cor_fun}c. 
The correlation function shows a clear anisotropic pattern which is 
essentially symmetric about $\Delta\phi = 90^o$. Such a pattern is consistent 
with the features expected for in-plane elliptic flow, and serve to confirm 
the utility of azimuthal correlation functions for flow measurements at RHIC. 
The solid curve in Fig.~\ref{fig:cor_fun} represents a fit to the correlation
function following Eq.~\ref{dist}; it provides a measure of the 
magnitude of the elliptic flow via the Fourier 
coefficient $\sqrt{{\lambda}_2} = v_2$.  

An important prerequisite for reliable flow extraction from PHENIX data
is to establish whether or not the $\sim 180^o$ azimuthal coverage of
the detector results in significant distortions to the 
correlation function. To this end, detailed simulations which take account
of the detector response, acceptance and efficiency have been performed
for simulated data in which specific amounts of elliptic flow were introduced. 
The essentially symmetric character of the correlation functions obtained from these simulations, provides a qualitative indication of the absence of distortions which could influence the 
reliability of the extracted flow. More quantitative
evidence have been provided by Fourier fits to the correlation functions; they indicate essentially complete recovery ($\sim 90\%$) of the input $v_2$ signals.  

\begin{figure}[htb]

\begin{minipage}[htb]{16cm}
\epsfig{file=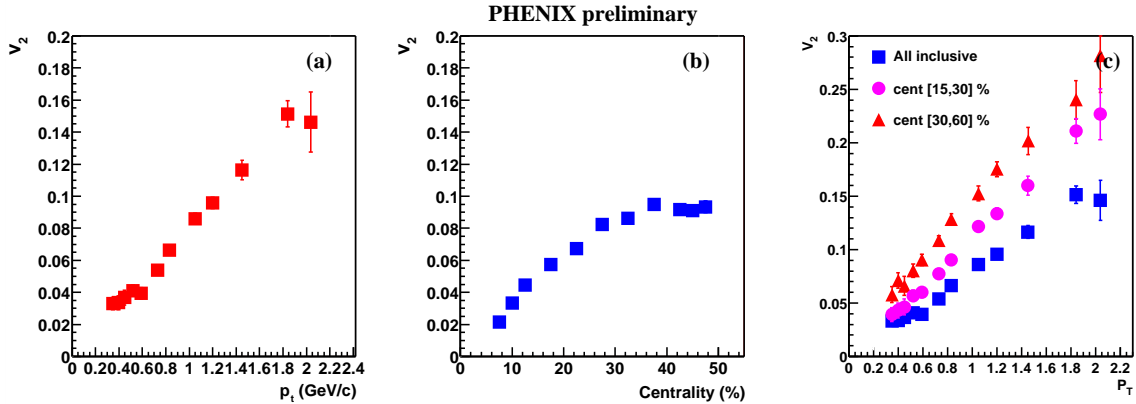,width=15.5cm}
\vskip -.75cm
\caption{\small \em $v_2$ vs. $p_t$ (a), $v_2$ vs. centrality (b), and $v_2$ vs. $p_t$
for several different centralities (inclusive, 15-30\%, and 30 - 60\% ) as indicated. 
Error bars indicate statistical errors only.}
\label{fig:pt_cent}
\end{minipage}
\end{figure}
\vskip -.5cm

\subsection{Differential Elliptic Flow}

   The magnitude of the elliptic flow and the mechanism for its development are 
related to several aspects: (a) the geometry of the collision zone, (b) the initial baryon and energy density developed in this zone, and (c) the detailed nature of the equation of state for the created nuclear matter\cite{teaney2001,kolb2001}. 
To disentangle these separate effects will undoubtedly require detailed differential elliptic flow studies. That is, $v_2(p_t)$, $v_2(cent)$, $v_2(\eta)$, etc. Initial 
differential flow measurements ($v_2(p_t)$, $v_2(cent)$, $v_2(p_t, cent)$) obtained with the PHENIX detector are summarized in Figs.~\ref{fig:pt_cent}; they represent the results obtained from azimuthal correlation functions for charged hadrons ($0.3 < p_t <2.5$ GeV/c) emitted in collisions at several centralities.   
Figs.~\ref{fig:pt_cent}a and b show the differential flow results, $v_2(p_t)$ and $v_2(cent)$ respectively. Fig.~\ref{fig:pt_cent}c 
compares the differential flow, $v_2(p_t, cent)$ for several centralities as indicated. Figs.~\ref{fig:pt_cent}a~-~c show relatively large differential flow values which increase with the $<p_t>$ of emitted hadrons, as well as with
increasing impact parameter. These magnitudes and trends are not only consistent with 
the results of other flow measurements at RHIC\cite{star_flow2000,talks}, but are in surprisingly good qualitative agreement with hydrodynamic model calculations\cite{teaney2001,kolb2001}. 
Since these calculations assume local thermal equilibrium at every space-time 
point in the collision zone, they are suggestive of the possibility that rapid thermalization occurs in Au~+Au collisions at RHIC. Such a notion is of course 
important to the task of delineating the EOS and establishing whether or not 
QGP formation occurs at RHIC.

\section{Summary}

	In summary, elliptic flow measurements have been made with the PHENIX detector 
via two-particle azimuthal correlation functions. The measurements indicate relatively large flow magnitudes (differential and integral). The 
differential flow $v_2(p_t)$, $v_2(cent)$ and $v_2(p_t, cent)$ are found to increase with both the 
$<p_t>$ of emitted hadrons, and with increasing impact parameter. 
The magnitude and the data trends are in good qualitative agreement with 
other flow measurements at RHIC\cite{talks}. They also show surprisingly 
good agreement with the results from hydrodynamic model calculations\cite{teaney2001,kolb2001}. Further detailed analyses of 
PHENIX flow data are currently in progress and will be reported elsewhere\cite{phenix_cor}.

\end{document}